\documentclass{tlp}
\submitted{30 March 2003}
\revised{20 May 2004, 6 Januari 2005}
\accepted{30 May 2005}

\input epsf

\begin{document}

\long\def\comment#1{}

\title{Programming Finite-Domain Constraint Propagators in Action Rules}

\author[N.F. Zhou]
{Neng-Fa Zhou \\
Department of Computer and Information Science \\
CUNY Brooklyn College \& Graduate Center \\
zhou@sci.brooklyn.cuny.edu 
}

\pagerange{\pageref{firstpage}--\pageref{lastpage}}
\volume{\textbf{10} (3):}
\jdate{March 2002}
\setcounter{page}{1}
\pubyear{2002}

\maketitle

\label{firstpage}

\begin{abstract}
In this paper, we propose a new language, called AR ({\it Action Rules}), and describe how various propagators for finite-domain constraints can be implemented in it. An action rule specifies a pattern for agents, an action that the agents can carry out, and an event pattern for events that can activate the agents. AR combines the goal-oriented execution model of logic programming with the event-driven execution model. This hybrid execution model facilitates programming constraint propagators. A propagator for a constraint is an agent that maintains the consistency of the constraint and is activated by the updates of the domain variables in the constraint. AR has a much stronger descriptive power than {\it indexicals}, the language widely used in the current finite-domain constraint systems, and is flexible for implementing not only interval-consistency but also arc-consistency algorithms. As examples, we present a weak arc-consistency propagator for the {\tt all\_distinct} constraint and a hybrid algorithm for n-ary linear equality constraints. B-Prolog has been extended to accommodate action rules. Benchmarking shows that B-Prolog as a CLP(FD) system significantly outperforms other CLP(FD) systems.
\end{abstract}

\section{Introduction}
CLP(FD), the constraint logic programming language over finite domains, has been proved effective for solving a large number of real-life optimization problems \cite{DIN90,Jaffar94}. The key operation employed in CLP(FD) is called {\it constraint propagation} \cite{KUM92,TSA93}, which uses constraints actively to prune search spaces as follows: whenever a variable changes, i.e., the variable has been instantiated or its domain has been updated, the domains of all the remaining variables are  filtered to contain only those values that are consistent with this variable. There may exist different propagation rules for a constraint depending on the level of consistency to be achieved. Constraint propagation has been employed to solve not only constraints over finite domains but also constraints over trees, lists, finite sets, floating-point numbers, and many other domains \cite{Jaffar94}.

In early CLP(FD) systems, such as the CHIP system \cite{DIN88}, constraints are interpreted rather than compiled. Constraints are first transformed into canonical-form terms and are then executed by an interpreter that performs, among other things, constraint propagation. The propagation procedure adopted is general enough for handling all types of constraints. Learning from the success of compiling Prolog programs into the Warren Abstract Machine (WAM) \cite{Warren83}, a former research group at ECRC extended the WAM for compiling CLP(FD) \cite{AGB91}. The CHIP compiler compiles constraints into low-level instructions such that different specialized propagation procedures are used for different types of constraints. This black-box approach has proved problematic because it is too complicated and lacks flexibility and extendibility. The extended WAM in the CHIP system \cite{AGB91} has over 100 instructions for compiling finite-domain constraints alone!

A language construct, called {\it indexicals}, has been quite popular as an intermediate language for compiling finite-domain constraints. The language was first proposed in \cite{HEN92} and then popularized by \cite{COD96}. This language is also adopted by other systems \cite{Carlsson97,SID96}. An indexical is a primitive constraint in the form of {\it X in r}, where $X$ is a domain variable and $r$ is a range expression for $X$. For each indexical, a propagation procedure specific to it is used. Indexicals are claimed to be a glass-box approach to compiling constraints in contrast with the black-box CHIP compiler. Nevertheless, as the delaying mechanism is embedded in range expressions, indexicals are not as open as claimed. Indexicals can be used to compile arithmetic constraints, but are too weak to be used to program many other kinds of propagators.

CHR (Constraint Handling Rules) \cite{FRU98} may currently be the most powerful implementation language for constraints. It can be used to program not only constraint propagators but also constraint reasoning rules. CHR has been implemented and integrated with ECLiPSe, SICStus, HAL, and Oz. CHR resembles a production system. In CHR, the left-hand side of a rule specifies a pattern of constraints in the constraint store and the right-hand side specifies new constraints to replace those on the left-hand side or to be added into the store. The left-hand side of a rule may have multiple constraint patterns. This feature is helpful for reasoning about the constraint store. For example, $A > B\ \&\ B > C \rightarrow A > C+1$ is a CHR rule that generates the constraint $A > C+1$, which is helpful albeit redundant. The strong descriptive power, however, is not offered without cost. For CHR, a sophisticated matching algorithm is needed to match constraint patterns against the constraint store. Now, constraint solvers implemented in CHR are still an order of magnitude slower than constraint interpreters implemented in C \cite{HOL99,HOL04}.

This paper proposes a new language, called AR ({\it Action Rules}), which can be used to program event handling in general and constraint propagation in particular. An action rule specifies a pattern for agents, an action that the agents can carry out, and an event pattern for events that can activate the agents. An agent behaves in an event-driven manner. An agent can be suspended when certain conditions on it are satisfied and can be activated when certain events are posted. AR is an extension of delay constructs such as delay clauses \cite{Meier94} that allows for the descriptions of not only delay conditions but also activating events and actions \cite{Zhou98}. The syntax, operational semantics, and implementation of AR will be described in Section 3.

The focus of this paper is on how to implement various propagators for finite-domain constraints in AR. A propagator for a constraint is an agent that maintains the consistency of the constraint and is activated by the updates of the domain variables in the constraint. In Section 4, we present propagators for binary, non-binary, and the global constraint {\tt all\_distinct}. AR is more expressive than indexicals. Some of the propagators presented, such as the one for maintaining arc consistency for binary equality constraints and the one for maintaining weak arc consistency for {\tt all\_distinct}, cannot be implemented in indexicals as efficiently.

B-Prolog has been extended to accommodate AR and several constraint solvers including the ones over finite domains, Boolean, trees, and finite sets have been developed in AR \cite{Zhou02}. Section 5 compares the performance of the finite-domain solver of B-Prolog with GNU-Prolog (GP), a state-of-the-art implementation of CLP(FD) \cite{Diaz01}, and two other CLP(FD) systems: ECLiPSe and SICStus. Benchmarking results show that B-Prolog is significantly faster than GP and 4-6 times as fast as ECLiPSe and SICStus.

Readers are assumed to be familiar with logic programming and constraint satisfaction, but no knowledge about the compilation is assumed. In Section 2, we define some preliminary terms and concepts about CLP(FD) and constraint propagation. Readers are referred to \cite{MAR98}, \cite{HEN89} and \cite{KUM92} for the details. A brief description of the implementation of AR is given in Section 3, and a more detailed description can be found in \cite{Zhou03tr}. 

\section{Preliminaries}
\subsection{CLP(FD)}
CLP(FD) \cite{HEN89} is an extension of Prolog that supports built-ins for specifying domain variables, constraints, and strategies for instantiating variables. 

The domains of variables are declared as follows: \[Vars\ {\tt in}\ D\]

\noindent
where $Vars$ is a variable or a list of variables, and $D$ is a list of ground terms or a range between two integers $l$..$u$. A domain variable is normally represented as a Prolog variable with attributes. A CLP(FD) system provides primitives for accessing and updating attribute values.

A CLP(FD) system provides {\it equality} ($=$), {\it disequality} ($\neq$), and {\it inequality} constraints. In addition, a CLP(FD) system also provides some other constraints such as global constraints. The global constraint {\tt all\_distinct(}$L${\tt )} ensures that the elements in the list $L$ must be all pairwise different. 

\subsection{Constraint propagation}
{\it Constraint Propagation} \cite{KUM92,TSA93} is a key operation employed in CLP(FD) systems for maintaining the consistency of constraints. The basic idea of constraint propagation is to activate the propagators of constraints whenever the domains of the variables in the constraints are updated. Propagating the updates to other variables may result in the shrinking of the domains of the variables or the instantiation of the variables.

There are different levels of consistency for constraints such as {\it node}, {\it interval}, {\it bounds}, {\it arc}, and {\it path} consistency \cite{TSA93,MAR98}. We define below three levels of consistency needed in this paper, namely node, interval and arc consistency, and define the propagators that maintain them. 

A unary constraint $p(X)$, where {\it X} has the domain {\it D}, is said to be {\it node-consistent} if, for any element {\it x} in {\it D}, $p(x)$ is satisfied. 
\[\forall_{x \in D}p(x)\]
For example, for the equality constraint $X = Y+1$, when $X$ is instantiated to $3$, the constraint becomes unary and $Y$ must be instantiated to $2$ to make the constraint node-consistent. As another example, for the disequality constraint $X \neq Y$, when $X$ is instantiated to $3$, $3$ must be excluded from the domain of $Y$ to make the constraint node-consistent. The propagation rule that maintains node consistency is called {\it forward checking}. A propagator for a constraint that performs forward checking is activated whenever the constraint becomes unary. 

Let $C$ be a linear equality constraint $c+a_1\times X_1+a_2\times X_2+\ldots+a_n\times X_n=0$ where $a_i \neq 0$ and $X_i$ is defined over the domain $D_i$ ($i=1,\ldots,n$). Let \[g_i(X_1,\ldots,X_{i-1},X_{i+1},\ldots,X_n)=\] 
\[\frac{-c-a_1\times X_1-\ldots-a_{i-1}\times X_{i-1} - a_{i+1}\times X_{i+1}-\ldots-a_n\times X_n}{a_i}\] and $l$ and $u$ be the functions defined as follows:
\[l(g_i(X_1,\ldots,X_{i-1},X_{i+1},\ldots,X_n)) = \] 
\[min\{g_i(x_1,\ldots,x_{i-1},x_{i+1},\ldots,x_n) | x_k \in D_k, 1\le k \le n, k \neq i\} \]
\[u(g_i(X_1,\ldots,X_{i-1},X_{i+1},\ldots,X_n)) = \] 
\[max\{g_i(x_1,\ldots,x_{i-1},x_{i+1},\ldots,x_n) | x_k \in D_k, 1\le k \le n, k \neq i\} \]
The constraint $C$ is said to be {\it interval consistent w.r.t. $X_i$} if: \\

\noindent
$\forall_{x\in D_i}(l(g_i(X_1,\ldots,X_{i-1},X_{i+1},\ldots,X_n)) \le x \le u(g_i(X_1,\ldots,X_{i-1},X_{i+1},\ldots,X_n))$ \\

To make the constraint interval consistent w.r.t. $X_i$, we have to exclude all the elements from $D_i$ that are not in the range. The constraint $C$ is said to be {\it interval consistent} if $C$ is interval consistent w.r.t. all the variables. For example, the constraint $X = Y+1$, where $X$ and $Y$ have the domain $1$..$5$, is not interval-consistent. To make the constraint interval consistent, we have to exclude $1$ from the domain of $X$ and $5$ from the domain of $Y$. Propagators for maintaining interval consistency are activated whenever a bound of a variable is updated or whenever a variable is instantiated. The definition can be easily extended to an inequality constraint.

Consider a binary constraint $p(X,Y)$ where $X$ and $Y$ are defined over the domains $D_x$ and $D_y$, respectively. The constraint is said to be {\it arc-consistent w.r.t. $X$} if for any element in $D_x$ there exists a supporting element in $D_y$ such that the constraint is satisfied:
\[\forall_{x \in D_x}\exists_{y \in D_y} p(x,y)\]
Similarly, the constraint is arc-consistent w.r.t. $Y$ if for any element in $D_y$ there exists a supporting element in $D_x$ such that the constraint is satisfied:
\[\forall_{y \in D_y}\exists_{x \in D_x} p(x,y)\]
The constraint is {\it arc-consistent} if it is arc-consistent w.r.t. both $X$ and $Y$. For example, the equality constraint $X = Y+1$ ($X \in \{2,4,5\}$, $Y \in \{1$..$4\}$) is not arc-consistent since there is no element in the domain of $X$ that supports $2$ in the domain of $Y$. To make the constraint arc-consistent, we must exclude $2$ from the domain of $Y$. Propagators for maintaining arc consistency are triggered whenever changes occur to the domain of a variable. Maintaining arc consistency for a non-binary constraint requires examining Cartesian products of the domains \cite{Dechter03} and is thus very costly. For this reason, some CLP(FD) systems maintain arc consistency only for binary constraints and many others do not consider arc consistency at all.

\subsection{Domain variables}
\label{sec_domain_variables}
A {\it domain variable} is a suspension variable to which there are suspended propagators and some other information attached.
A domain variable is represented in B-Prolog as a record that has the following fields:
\begin{center}
\begin{tabular}{ll} \hline
$ref$ & reference to the value \\
$type$ & type of the domain \\
$min$ & minimum element in the domain\\
$max$ & maximum element in the domain\\
$size$ & number of elements that remain in the domain \\
$ins\_cs$ & list of propagators to be executed when the variable is instantiated \\
$bound\_cs$ & list of propagators to be executed when a bound is updated \\
$dom\_cs$ & list of propagators to be executed when an inner element is excluded \\ 
$elms$ & pointer to the bit vector representation of the elements \\ \hline
\end{tabular}
\end{center}
where $ref$ refers to the variable itself if the variable is not instantiated, and $elms$ is a pointer to a bit vector that represents the elements. When a domain is an interval without holes, no bit vector is necessary and $elms$ is a null pointer.

The following built-in predicates and functions are available on domain variables. 
\begin{itemize}
\item {\tt dvar(}$X${\tt )}: Succeeds if $X$ is a domain variable.
\item {\tt min(}$X${\tt )}, {\tt max(}$X${\tt )}: Functions that return, respectively, the minimum and maximum elements of the domain of $X$.
\item {\tt size(}$X,Size${\tt )}: The size of the domain of $X$ is $Size$.
\item {\tt exclude(}$X,E${\tt )}: Excludes the element $E$ from the domain of $X$.
\item $X${\tt in}$\ D$: The new domain for $X$ is the intersection of its existing domain and $D$ where $D$ is a set of ground terms or a range $l$..$u$ of integers.
\end{itemize}
A failure occurs when the domain of a variable becomes empty. When the domain of a variable becomes a singleton, the variable is instantiated to the element automatically.

An event is posted whenever the domain of a variable is updated. For a domain variable $X$, instantiating $X$ posts the event {\tt ins(}$X${\tt )},\footnote{A variable is said to be instantiated if it is bound to another term, possibly another variable.} updating the lower or upper bound of the domain posts the event {\tt bound(}$X${\tt )}, and excluding an inner element $E$ from the domain posts the event {\tt dom(}$X,E${\tt )}. Notice that the event {\tt bound(}$X${\tt )} is not posted when $X$ is instantiated and the event {\tt dom(}$X,E${\tt )} is not posted when either bound of the domain of $X$ is updated. This implies that a propagator that maintains arc consistency has to handle not only {\tt dom(}$X,E${\tt )} events but also {\tt bound(}$X${\tt )} and {\tt ins(}$X${\tt )} events.

Each event on a domain variable activates its corresponding list of propagators. The event {\tt ins(}$X${\tt )} activates the propagator list $ins\_cs$ of $X$, {\tt bound(}$X${\tt )} activates the list $bound\_cs$, and {\tt dom(}$X,E${\tt )} activates the list $dom\_cs$. 

\section{The AR Language}
AR is designed for programming interactive agents. In this section, we describe the syntax, operational semantics, and implementation of AR. 

\subsection{Syntax}
An {\it action rule} takes the following form: \[Agent, Condition, \{Event\} => Action\]
where $Agent$ is an atomic formula that represents a pattern for agents, $Condition$ is a conjunction of conditions on the agents, $Event$ is a non-empty disjunction of patterns for events that can activate the agents, and $Action$ is a sequence of subgoals.\footnote{Subgoals in $Action$ can be any subgoals including those defined by Prolog clauses.} $Condition$ and the following comma can be omitted if no condition is needed on $Agent$. $Action$ cannot be empty. The subgoal {\tt true} represents an empty action that always succeeds. An action rule degenerates into a {\it commitment rule} if $Event$ together with the enclosing braces are missing. Conditions, event patterns, and actions are all atomic formulas where the delimiter `,' is used to separate the constituents. 

An AR {\it predicate} consists of a sequence of rules defining agents of the same predicate symbol. In a program, AR predicates can be intermingled with Prolog predicates. In this paper, the term {\it agents} is used to refer to subgoals that can be suspended and activated, and the term {\it predicate} is used refer to both AR and Prolog predicates unless explicitly specified.

All conditions must be {\it in-line tests}.\footnote{An in-line call is compiled into instructions that do not invoke any predicates. A test does not change the instantiation status of the variables in its arguments.} In the implementation of AR in B-Prolog, the following types of conditions are allowed:

\begin{itemize} 
  \item Type and mode checking: Predicates like {\tt integer(}$X${\tt )}, {\tt var(}$X${\tt )}, and {\tt nonvar(}$X${\tt )}.

  \item Matching: A matching call takes the form of $X=Y$ where one of the arguments must be a non-variable term at compile time and the other must be a variable (again at compile time) that occurs before in the rule. The non-variable term serves as a pattern and the variable refers to a term to be matched against the pattern. This call succeeds if the pattern and the term become identical after a substitution is applied to the pattern. For instance, the condition $f(X)=Y$ succeeds if $Y$ is a structure whose functor is $f/1$.

  \item Term inspection: Several built-in predicates including {\tt arg/3}, {\tt functor/3}, {\tt $\dequals$/2}, {\tt $\setminus\dequals$/2}, and {\tt n\_vars\_gt/2} can be used in the condition of a rule to inspect the arguments of an agent. The call {\tt n\_vars\_gt(}$Term,N${\tt )} succeeds if the number of variables occurring in $Term$ is greater than $N$. 

\item Arithmetic comparison: Checks the arithmetic equality ({\tt $=$$:$$=$}), disequality ({\tt $=$$\setminus$$=$}), or inequality ({\tt $>$}, {\tt $>$$=$}, {\tt $<$}, and {\tt $=<$}) of two terms which must be ground at runtime.
\end{itemize}

A set of built-in events is provided.\footnote{In the implementation in B-Prolog, built-in events are provided for programming constraint propagators, graphical user interfaces, and interactive agents.} As far as programming constraint propagators is concerned, an event pattern can be one of the following: 
\begin{itemize}
\item {\tt generated}: The action of the rule is executed when the agent is suspended for the first time.
\item {\tt ins(}$X${\tt )}: The agent is activated when an event {\tt ins(}$X${\tt )} is posted.
\item {\tt bound(}$X${\tt )}: The agent is activated when an event {\tt bound(}$X${\tt )} is posted.
\item {\tt dom(}$X${\tt )} and {\tt dom(}$X,E${\tt )}: The agent is activated when an event {\tt dom(}$X,E'${\tt )} is posted. Before the action is executed, $E$ is made to reference the element $E'$.
\end{itemize}

A user program can create and post its own events and define agents to handle them. A user-defined event takes the form of {\tt event(}$X,T${\tt )} where $X$ is a suspension variable that connects the event with its handling agents, and $T$ is a Prolog term that contains the information to be transmitted to the agents. If the event poster does not have any information to be transmitted to the agents, then the second argument $T$ can be omitted. The built-in {\tt post(}$E${\tt )} posts the event $E$.

In an action rule, the event pattern {\tt dom(}$X,T${\tt )} or {\tt event(}$X,T${\tt )} is not allowed to coexist with any other event patterns and $T$ must be a first-occurring variable so that when the action of the rule is executed $T$ always refers to the second argument of the event.

\subsection{Examples}
The following defines an agent that echoes the messages sent to it by event posters. 
\begin{verbatim}
      echo_agent(X), {event(X,Message)} => write(Message).
\end{verbatim}
The following query,
\begin{verbatim}
      echo_agent(Ping), echo_agent(Pong), 
      post(event(Ping,ping)), post(event(Pong,pong))
\end{verbatim}
generates two echo agents {\tt echo\_agent(Ping)} and {\tt echo\_agent(Pong)}, and activates them by posting two events. The event {\tt event(Ping,ping)} activates the agent {\tt echo\_agent(Ping)}, and the event {\tt event(Pong,pong)} activates {\tt echo\_agent(Pong)}.

The following defines the freeze predicate in Prolog-II \cite{Colmerauer84}. 
\begin{verbatim}
    freeze(X,G), var(X), {ins(X)} => true.
    freeze(X,G) => call(G).
\end{verbatim}
The primitive {\tt freeze(X,G)} is logically equivalent to {\tt call(G)} but the execution of {\tt G} is delayed until {\tt X} is instantiated to a non-variable term. The agent {\tt freeze(X,G)} is suspended waiting for an event {\tt ins(X)} when {\tt X} is a variable. When an event {\tt ins(X)} is posted, the condition {\tt var(X)} is tested {\it again}. If it succeeds, then the action {\tt true} is executed and the agent becomes suspended again. As long as {\tt X} is a free variable, the agent {\tt freeze(X,G)} is suspended. Only when {\tt X} becomes a non-variable term, can the second rule be applied.

Consider, as another example, how to implement the following indexical:
\begin{verbatim}
    X in min(Y)+min(Z)..max(Y)+max(Z).
\end{verbatim}
which ensures that the constraint {\tt X = Y+Z} is interval-consistent w.r.t. {\tt X}. 
\begin{verbatim}
    `V in V+V'(X,Y,Z),{generated,ins(Y),bound(Y),ins(Z),bound(Z)} =>
          reduce_domain(X,Y,Z).

    reduce_domain(X,Y,Z) =>
          L is min(Y)+min(Z), U is max(Y)+max(Z),
          X in L..U.
\end{verbatim}
The propagator is activated whenever a bound of {\tt Y} or {\tt Z} is updated or either one is instantiated. The action {\tt reduce\_domain(X,Y,Z)} enforces that the domain of {\tt X} be in the range {\tt min(Y)+min(Z)..max(Y)+max(Z)}. The action is also executed before the propagator is first suspended so that no preprocessing is needed to enforce interval consistency.
\begin{center}
\begin{figure}
\epsfxsize=8cm 
\centering{\epsfbox{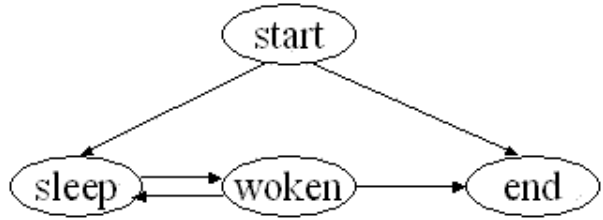}}
\caption{\label{fig:states}Diagram of state transition of agents.}
\end{figure}
\end{center}

\subsection{Operational Semantics}
The operational semantics of AR can be presented as a state-transition system as shown in Figure \ref{fig:states}. An agent may be in one of the following states: {\it start}, {\it sleep}, {\it woken}, and {\it end}. When an agent is generated, it enters the {\it start} state and is executed immediately. A typical agent transits to the {\it end} state through the {\it sleep} and {\it woken} states. An agent is said to be {\it floundering} if it stays in the {\it sleep} state forever. It is the programmer's responsibility to prevent agents from floundering.

When an agent is generated, the system searches in its predicate in textual order for a rule whose agent-pattern {\it matches} the agent and whose condition is satisfied. This kind of rule is said to be {\it applicable} to the agent. Formally, an action rule ``$H,C,\{E\} => B$'' or a commitment rule ``$H,C => B$'' is applicable to an agent $\alpha$ if there exists a unifier $\theta$ such that $H\theta=\alpha$ and $C\theta$ is satisfied.\footnote{Notice that since one-directional matching rather than full unification is used to search for an applicable rule and in the condition no variable in $\alpha$ can be instantiated, the agent remains the same after an applicable rule is found.} If no rule is found applicable, the agent fails. If a commitment rule is found, the agent is substituted for the body,\footnote{A commitment rule is similar to a guarded clause in concurrent logic languages \cite{SHA89}, but an agent can never be blocked while it is being matched against an agent pattern.} and its state is changed to {\it end}. 

If an action rule $``H,C,\{E\} => B$'' is found for the agent, the agent is  suspended, transiting from {\it start} to {\it sleep}. It will stay in the {\it sleep} state until it is activated by one of the events in $E$. 

When an event is posted, all the sleeping agents waiting for the event in the system are woken up and the event is erased after that so that no agents generated later will be responsive to this event. The woken agents are added to the queue of active subgoals in some order. It is up to the implementer of the language to use a strategy to schedule activated agents. Whatever scheduling strategy is adopted, the programmer should not rely on the strategy to guarantee the correctness of programs. 

Suppose an agent was put into {\it sleep} by the action rule $``H,C,\{E\} => B$'' and was woken up by one of the events in $E$. After this agent is picked by the scheduler, the system tests the condition  $C$ again. If it is satisfied, the action $B$ is executed. If the action succeeds, the agent is re-suspended. If the action fails, the agent fails as well. If the condition $C$ of the action rule does not hold, the system searches for an alternative applicable rule for the agent just as for a newly generated agent.

There is no primitive for killing agents explicitly. An agent never disappears as long as action rules are applied to it successfully. An agent transits to the {\it end} state only when a commitment rule is applied to it. 

\subsection{The Implementation}
The abstract machine of B-Prolog, called ATOAM \cite{Zhou96}, is extended to support agents \cite{Zhou03tr}. This subsection briefly describes this implementation. A more detailed description can be found in \cite{Zhou03tr}.

\subsubsection{The frame structure for agents}
The ATOAM is a variant of the Warren Abstract Machine (WAM) \cite{Warren83}. Unlike in the WAM where arguments are passed through argument registers, arguments in the ATOAM are passed through stack frames and only one frame is used for each subgoal. Each time a predicate is invoked by a subgoal, a frame is placed on top of the control stack unless the frame currently at the top can be reused. Frames for different types of predicates have different structures. 

Agents are stored as frames on the control stack. The frame for an agent has the following slots in addition to those included in a normal frame:\footnote{A normal frame has the following slots: arguments, FP (parent frame), CPS (continuation program pointer on success), TOP (top of the control stack), BTM (bottom of the frame), and local variables. A choice point frame has the following additional slots: CPF (continuation program pointer on failure),  B (parent choice point),  H (top of the heap), and T (top of the trail stack).} 
\begin{center}
\begin{tabular}{ll} 
STATE: &     State of the agent \\
EVENT: &     Activating event \\
REEP: &     Re-entrance program pointer \\
PREV: &     Previous agent in the chain \\
\end{tabular}
\end{center}
The STATE slot has one of those states shown in Figure \ref{fig:states} as its value. The EVENT slot stores the last event that activated the agent. The action rule of the agent can have access to this event. The REEP slot stores the program pointer to continue when the agent is activated. The PREV slot stores the pointer to the previous agent's frame in the chain of agents.

The frames on the control stack comprise three chains, namely the chain of {\it active subgoals}\footnote{i.e., the frames connected by FP.} that are being executed, the chain of {\it choice points}, i.e., subgoals that have alternative clauses to be tried when execution backtracks to them, and the chain of {\it agents} in either {\it sleep} or {\it woken} states. 

Storing agents on the stack facilitates context switching for agents \cite{Zhou961} but complicates memory management. With frames of agents on the stack, the chronological order of frames is no longer preserved, and therefore a garbage collector is needed to collect useless frames on the control stack and run-time checking is needed to determine whether the current frame can be reused.

\subsubsection{When and how to invoke agents?}
For the sake of efficiency, events are not checked after each instruction but checked at the entry and exit points of each predicate. If it is found that the list of events is not empty, those agents that are waiting for the events are added into the active chain and the current predicate is interrupted. After the activated agents complete their execution successfully, the interrupted predicate resumes its execution. The programmer has no control over the order in which agents are added. In our implementation, the {\it first-generated-first-served} strategy is used.

At a point during execution, there may be multiple events posted that are all expected by an agent. If this is the case, then the agent must be activated once for each of the events. If an agent is found to be active already when the system tries to add it into the active chain, the sytem makes a copy of it and adds the copy into the chain. 

The actions of constraint propagators are to reduce the domains of variables. This characteristic is exploited to improve the performance of constraint propagators. Some events that cannot lead to the shrinking of any domains are ignored. For example, if multiple events of {\tt bound(X)} are posted at the same time, then only one of them needs to be handled, and if {\tt bound(X)} and {\tt ins(X)} are posted at the same time, then the {\tt bound(X)} event is ignored. In this way, many redundant activations of rules that do not contribute to the reduction of any domains can be suppressed. This optimization is applied to constraint propagators only and not general agents.

In a CLP(FD) program, constraint propagation is normally intertwined with non-deterministic subgoals such as {\tt labeling} that assign values to variables. If a non-deterministic predicate is interrupted by events, then no choice point can be created for the interrupted predicate until the activated agents are all executed. Consider the following example:
\begin{verbatim}
      ?-p(X),X=f(Y),q(X),write(X).

      p(X),var(X),{ins(X)} => true.
      p(X) => X=f(a).

      q(X):-fail.
      q(X).
\end{verbatim}
First the agent {\tt p(X)} is generated, waiting for {\tt X} to be instantiated. The subgoal {\tt X=f(Y)} posts an event {\tt ins(X)} after {\tt X} bound to {\tt f(Y)}. At the entry of {\tt q(X)}, the event is detected and {\tt p(X)} is activated. If there were a choice point created for {\tt q(X)} before {\tt p(X)} is activated, the binding {\tt Y=a} given by the second rule of {\tt p(X)} would be lost when {\tt fail} in {\tt q/1} is executed since {\tt Y} is older than the choice point, and the output from {\tt write(X)} would be {\tt f(Y)} not {\tt f(a)}.

\section{Programming Constraint Propagators in AR}
The high descriptive power of AR opens new ways to implementing constraint propagators. In this section, we implement propagators that maintain node, interval, and arc consistency for binary constraints, a hybrid algorithm for non-binary constraints, and a weak arc-consistency propagator for the global constraint {\tt all\_distinct}.

\subsection{Binary constraints}
We consider how to implement propagators for the binary constraint $A\times X = B\times Y+C$, where $X$ and $Y$ are domain variables, $A$ and $B$ are positive integers, and $C$ is an integer of any kind. Similar propagators can be implemented for other types of binary constraints.

\subsubsection{Forward checking}
Recall that forward checking enforces node consistency. The following shows a propagator that performs forward checking for the binary constraint. 
\begin{verbatim}
    `aX=bY+c'(A,X,B,Y,C) =>
          `aX=bY+c_forward'(A,X,B,Y,C).

    `aX=bY+c_forward'(A,X,B,Y,C),var(X),var(Y),{ins(X),ins(Y)} => true.
    `aX=bY+c_forward'(A,X,B,Y,C),var(X) => 
           T is B*Y+C, X is T//A, A*X=:=T.
    `aX=bY+c_forward'(A,X,B,Y,C) => 
           T is A*X-C, Y is T//B, B*Y=:=T.
\end{verbatim}
The operation {\tt op1//op2}, which is equivalent to {\tt truncate(op1/op2)}, gives the integer quotent of the division. When both {\tt X} and {\tt Y} are variables, the propagator is suspended. When either variable is instantiated, the propagator computes the value for the other variable. 

\subsubsection{Interval consistency}
The following propagator, which extends the forward-checking propagator, maintains interval consistency for the constraint. 
\begin{verbatim}
    `aX=bY+c'(A,X,B,Y,C) =>
          `aX=bY+c_reduce_domain'(A,X,B,Y,C),
          `aX=bY+c_forward'(A,X,B,Y,C),
          `aX=bY+c_interval'(A,X,B,Y,C).
\end{verbatim}

The subgoal {\tt `aX=bY+c\_reduce\_domain'(A,X,B,Y,C)} preprocess the constraint to make it interval-consistent when the constraint is generated.
\begin{verbatim}
    `aX=bY+c_reduce_domain'(A,X,B,Y,C) =>
          `aX in bY+c_reduce_domain'(A,X,B,Y,C),
          MC is -C,
          `aX in bY+c_reduce_domain'(B,Y,A,X,MC).

    `aX in bY+c_reduce_domain'(A,X,B,Y,C) =>
          L is (B*min(Y)+C) /> A, 
          U is (B*max(Y)+C) /< A,
          X in L..U.
\end{verbatim}
The operation {\tt op1 /> op2} returns the lowest integer that is greater than or equal to the quotient of {\tt op1} by {\tt op2} and the operation {\tt op1 /< op2} returns the greatest integer that is less than or equal to the quotient. It can be proved easily that no value outside the range {\tt L..U} satisfies the constraint.

The subgoal {\tt `aX=bY+c\_interval'(A,X,B,Y,C)} maintains interval consistency for the constraint.
\begin{verbatim}
    `aX=bY+c_interval'(A,X,B,Y,C) =>
          `aX in bY+c_interval'(A,X,B,Y,C),  % reduce X when Y changes
          MC is -C,
          `aX in bY+c_interval'(B,Y,A,X,MC). % reduce Y when X changes
       
    `aX in bY+c_interval'(A,X,B,Y,C),
          var(X),var(Y),
          {generated,bound(Y)} 
          =>
          `aX in bY+c_reduce_domain'(A,X,B,Y,C).
    `aX in bY+c_interval'(A,X,B,Y,C) => true.
\end{verbatim}
Notice that the action {\tt `aX=bY+c\_reduce\_domain'(A,X,B,Y,C)} is executed only when both variables are free. If either one turns to be instantiated, then the  forward-checking rule takes care of that situation.

\subsubsection{Arc consistency}
The following propagator, which extends the one shown above, maintains arc consistency for the constraint.
\begin{verbatim}
    `aX=bY+c'(A,X,B,Y,C) =>
          `aX=bY+c_reduce_domain'(A,X,B,Y,C),
          `aX=bY+c_forward'(A,X,B,Y,C),
          `aX=bY+c_interval'(A,X,B,Y,C),
          `aX=bY+c_arc'(A,X,B,Y,C).

     `aX=bY+c_arc'(A,X,B,Y,C) =>
          `aX in bY+c_arc'(A,X,B,Y,C), % reduce X when Y changes
          MC is -C,
          `aX in bY+c_arc'(B,Y,A,X,MC).% reduce Y when X changes

     `aX in bY+c_arc'(A,X,B,Y,C),var(X),var(Y),{dom(Y,Ey)} => 
          T is B*Ey+C,
          Ex is T//A,
          (A*Ex=:=T -> exclude(X,Ex);true).
     `aX in bY+c_arc'(A,X,B,Y,C) => true.
\end{verbatim}
Whenever an element {\tt Ey} is excluded from the domain of {\tt Y}, the propagator {\tt `aX in bY+c\_arc'(A,X,B,Y,C)} is activated. If both {\tt X} and {\tt Y} are variables, the propagator excludes {\tt Ex}, the counterpart of {\tt Ey}, from the domain of {\tt X}. Again, if either {\tt X} or {\tt Y} becomes an integer, the propagator does nothing. The forward checking rule takes care of that situation. 

\subsection{Non-binary Constraints}
In indexical-based CLP(FD) systems, constraints are split into indexicals that contain no more than three variables. This algorithm has several advantages. Firstly, it generates linear-size code. Secondly, indexicals can be implemented in a low-level language to achieve better performance. Thirdly, information propagation can be restricted to only those constraints for which the domains have the possibility to be reduced \cite{COD96}. For example, consider the two ternary constraints $T1 = X1+X2$ and $T1+X3+X4 = 0$. If $T1 = X1+X2$ is activated by an update of $X1$, as long as the shared variable $T1$ does not change the other constraint needs not be activated. The disadvantages of this algorithm are that new domain variables have to be introduced and the granularity of constraints becomes smaller and thus context switching becomes more costly. In B-Prolog, each domain variable takes at least 10 words, letting alone the space for the constraints and data structures for the elements. The space overhead cannot be neglected when the number of variables is large.

In comparison with indexicals, the high descriptive power of AR opens new ways to compiling non-binary constraints. We present two algorithms. One is called {\it unite}, which adopts one propagator for each constraint to maintain interval consistency. The other one, called {\it hybrid}, maintains interval consistency when the constraint contains more than two variables and maintains arc consistency when the constraint turns into binary.

\subsubsection{Unite: use one propagator for each constraint}
Let $A_1\times X_1+\ldots+A_n\times X_n+C$ $=$ $0$ be an n-ary constraint where each $A_i$ ($i=1,\ldots,n$) is a non-zero integer and each $X_i$ is a domain variable or an integer. The propagator for the constraint takes the following form:
\begin{verbatim}
    `A1*X1+...+An*Xn+C=0'(C,A1,A2,...,An,X1,X2,..,Xn),
          {generated,ins(X1),bound(X1),...,ins(Xn),bound(Xn)}
    => 
          ... % reduce the domains of X1,...,Xn.
\end{verbatim}
In the action, attempts are made to reduce the lower and upper bounds of the domain of every variable.

To facilitate the generation of the code for reducing domains, the compiler splits the expression $A_1\times X_1+\ldots+A_n\times X_n+C$ into the following list of sub-expressions each of which contains at most three variables:

\begin{tabular}{l}
      $T_0 = C$, \\
      $T_1 = T_0+A_1*X_1$,\\
      $T_2 = T_1+A_2*X_2$, \\
      $\ldots$ \\
      $T_n = T_{n-1}+A_n*X_n$ \\
\end{tabular}

\noindent
The generated reducer first computes the lower and upper bounds of the temporary variables\footnote{Temporary variables are plain variables, not domain variables. Therefore, this compilation scheme is different from compiling constraints into indexicals.} by propagating information forward from $T_0$ to $T_n$. The lower and upper bounds of $T_i$ are computed from those of $T_{i-1}$ and $A_i\times X_i$ ($i=1,\ldots ,n$). After that, the reducer propagates information backward from $T_n$ to $T_1$. For each tuple $T_i = T_{i-1}+A_i\times X_i$, the new bounds of $T_{i-1}$ and $X_i$ are computed from those of $T_i$.

For example, the following shows the propagator generated for the constraint {\tt X1+X2+X3+C = 0}.
\begin{verbatim}
    `X1+X2+X3+C=0'(C,X1,X2,X3)
          {generated,ins(X1),bound(X1),ins(X2),bound(X2),
           ins(X3),bound(X3)}
    =>
          `X1+X2+X3+C=0_reducer'(C,X1,X2,X3).                    

    `X1+X2+X3+C=0_reducer'(C,X1,X2,X3) =>
          Lt1 is C+min(X1), Ut1 is C+max(X1),        % T1 = C+X1
          Lt2 is Lt1+min(X2), Ut2 is Ut1+max(X2),    % T2 = T1+X2
          Lt3 is Lt2+min(X3), Ut3 is Ut2+max(X3),    % T3 = T2+X3
          Lt3 =< 0, Ut3 >= 0,                        % T3 = 0
          %
          NewLx3 is 0-Ut2, NewUx3 is 0-Lt2,          % T3 = T2+X3
          X3 in NewLx3..NewUx3,
          NewLt2 is 0-max(X3), NewUt2 is 0-min(X3),
          %
          NewLx2 is NewLt2-Ut1, NewUx2 is NewUt2-Lt1,% T2 = T1+X2
          X2 in NewLx2..NewUx2,
          NewLt1 is NewLt2-max(X2), NewUt1 is NewUt2-min(X2),
          %
          NewLx1 is NewLt1-C, NewUx1 is NewUt1-C,    % T1 = C+X1
          X1 in NewLx1..NewUx1,
          NewLt1-max(X1) =< C, NewUt1-min(X1) >= C.
\end{verbatim}
The advantage of this algorithm is that only one propagator is used for each constraint whose code size is linear in the number of variables in the constraint. The weakness of this algorithm is that the reducer is not fast. Whenever a variable is instantiated or a variable's bound is updated, the reducer tries to reduce the domains of all the variables including the seed variable that triggers the propagator.

\subsubsection{Hybrid: combining interval and arc consistency algorithms}
For a non-binary constraint, it is too expensive to maintain arc consistency. One practical strategy is to maintain interval consistency while there are multiple variables in the constraint and to maintain arc consistency when the constraint turns into binary. The following shows the propagator for the linear non-binary constraint $A1\times X1+...+An\times Xn+C$$=$$0$.
\begin{verbatim}
    `A1*X1+...+An*Xn+C=0'(C,A1,A2,...,An,X1,X2,..,Xn),
          n_vars_gt([X1,...,Xn],2),
          {generated,ins(X1),bound(X1),...,ins(Xn),bound(Xn)} 
    => 
          ... % reduce domains of X1,...,Xn.
    `A1*X1+...+An*Xn+C=0'(C,A1,A2,...,An,X1,X2,..,Xn) =>
          nary_to_binary([C,A1,X2,A2,X2,...,An,Xn],NewC,B1,B2,Y1,Y2),
          call_binary_constraint_propagator(NewC,B1,Y1,B2,Y2).
\end{verbatim}
The propagator is activated whenever any variable is instantiated or its bound is updated. When {\tt n\_vars\_gt([X1,...,Xn],2)} succeeds, i.e. when there are multiple variables in the constraint, the domains are reduced to make the constraint interval-consistent. When the constraint becomes binary, the condition {\tt n\_vars\_gt} fails and the second rule is tried. The subgoal {\tt nary\_to\_binary} transforms the constraint into the binary constraint $B1\times Y1+B2\times Y2+NewC$$=$$0$, and the next subgoal invokes an appropriate propagator for the binary constraint.\footnote{In the implementation in B-Prolog, the two built-ins {\tt n\_vars\_gt} and {\tt nary\_to\_binary} do not take the constraint as an argument but instead access the constraint in the parent subgoal. In this way, no copy of the constraint needs to be made.}

\subsection{Propagators for {\tt all\_distinct}}
The constraint {\tt all\_distinct(}$L${\tt )} holds if the elements in $L$
are pairwise different. One naive implementation method for this
constraint is to generate binary disequality constraints between all
pairs of variables in $L$. This implementation has two problems:
First, the space required to store the constraints is quadratic in the
number of variables in $L$; Second, splitting the constraint into
fine-grained ones may lose possible propagation opportunities \cite{REG94,Puget98}. This subsection presents two propagators for the constraint. The propagation algorithms are not new. The goal of this subsection is to illustrate the expressive power of AR.

\subsubsection{A linear-space propagator}
To solve the space problem, we define {\tt all\_distinct} in the following way:
\begin{verbatim}
    all_distinct(L) => all_distinct(L,[]).

    all_distinct([],Left) => true.
    all_distinct([X|Right],Left) =>
          outof(X,Left,Right),
          all_distinct(Right,[X|Left]).

    outof(X,Left,Right), var(X), {ins(X)} => true.
    outof(X,Left,Right) => exclude_list(X,Left),exclude_list(X,Right).

    exclude_list(X,[]).
    exclude_list(X,[Y|Ys]):- exclude(Y,X),exclude_list(X,Ys).
\end{verbatim}
For each variable {\tt X}, let {\tt Left} be the list of variables to
the left of {\tt X} and {\tt Right} be the list of variables to the
right of {\tt X} in {\tt L}. The subgoal {\tt outof(X,Left,Right)} holds
if {\tt X} appears in neither {\tt Left} nor {\tt Right}. Instead of
generating disequality constraints between {\tt X} and all the
variables in {\tt Left} and {\tt Right}, the subgoal {\tt
outof(X,Left,Right)} suspends until {\tt X} is instantiated. After {\tt
X} becomes non-variable, {\tt exclude\_list(X,Left)} and {\tt
exclude\_list(X,Right)} exclude {\tt X} from the domains of the variables in
{\tt Left} and {\tt Right}, respectively. 

There is one propagator {\tt outof(X,Left,Right)} for each element {\tt X} in the list, which takes constant space. Therefore, {\tt all\_distinct(L)} takes linear space in the size of {\tt L}. Notice that the two lists {\tt Left} and {\tt Right} are not merged into one bigger list; otherwise, the constraint would take quadratic space.

\subsubsection{Weak arc consistency}
In terms of pruning ability, the linear-space propagator is the same as the naive one that splits a constraint of {\tt all\_distinct} into binary disequality constraints. In this subsection, we present a propagator that has stronger prunning power than the naive propagator.

Given any set of values $D$ of size $n$, the constraint {\tt all\_distinct(}$L${\tt )} is said to be {\it weak arc consistent} if there are at most $n$ variables in $L$ whose domains are subsets of $D$. For each variable $X$ in $L$, let $L-\{X\}$ be the list of variables in $L$ but $X$, $n$ be the size of the domain of $X$, and $m$ be the number of variables in $L-\{X\}$ whose domains are subsets of that of $X$. If $m+1>n$, then the constraint is unsatisfiable since it is impossible to assign $n$ values to more than $n$ variables such that each variable gets a different value. If $m+1=n$, then for each value $v$ in $X$'s domain, we can safely exclude $v$ from the domains of all the variables whose domains are not subsets of that of $X$. 

Consider the following query,
\begin{verbatim}
    X in {1,2}, Y in {1,2}, Z in {1,2}, all_distinct([X,Y,Z]).
\end{verbatim}
the weak arc-consistency propagator detects the inconsistency of the constraint without labeling any variables. For the following query,
\begin{verbatim}
    X in {1,2}, Y in {1,2}, Z in {1,2,3}, all_distinct([X,Y,Z]).
\end{verbatim}
the algorithm assigns {\tt 3} to {\tt Z} without labeling any variables.
    
The weak arc-consistency propagator is not as powerful as the algorithm proposed by Regin \cite{REG94} in terms of pruning ability but is much easier to implement. To incorporate weak arc-consistency checking into the linear-space propagator, we only need to redefine {\tt outof(X,Left,Right)} as follows:
\begin{verbatim}
    outof(X,Left,Right), var(X), {generated,ins(X),bound(X),dom(X)} =>
          outof_reducer(X,Left,Right).
    outof(X,Left,Right) => exclude_list(X,Left),exclude_list(X,Right).
\end{verbatim}
where {\tt outof\_reducer(X,Left,Right)} first counts the variables in {\tt Left} and {\tt Right} whose domains are subsets of the domain of {\tt X} and then decides what action to take depending on the count and the size of the domain of {\tt X}. 

The key operation is to decide whether a domain is a subset of another domain. In the worst case, the two domains have to be scanned. There are several facts that can be used to avoid scanning domain elements. A domain {\tt D1} cannot be a subset of another domain {\tt D2} if {\tt D1} has a larger size or has a larger interval. Also if two domains are intervals without holes, then scanning the elements is unnecessary. Another fact that can be used in the detection is that if the event is {\tt dom(X,E)} meaning that {\tt E} has been excluded from {\tt X}'s domain, then another domain {\tt Y} cannot be a subset of {\tt X} if {\tt E} is included in {\tt Y}. To take advantage of this fact, the propagator can be rewritten into the following:
\begin{verbatim}
    all_distinct(L) => all_distinct(L,[]).

    all_distinct([],Left) => true.
    all_distinct([X|Right],Left) =>
          outof(X,Left,Right),
          outof_dom(X,Left,Right),
          all_distinct(Right,[X|Left]).

    outof(X,Left,Right), var(X), {generated,ins(X),bound(X)} =>
          outof_reducer(X,Left,Right).
    outof(X,Left,Right) => exclude_list(X,Left),exclude_list(X,Right).

    outof_dom(X,Left,Right),var(X), {dom(X,E)} =>
          outof_reducer(X,E,Left,Right).
    outof_dom(X,Left,Right) => true.
\end{verbatim}
The subgoal {\tt outof\_reducer(X,E,Left,Right)} takes {\tt E} into account when detecting whether a domain is a subset of that of {\tt X}.

\section{Performance Evaluation}
B-Prolog has been extended to accommodate AR and the finite-domain constraint solver described in this paper has been developed in AR. In this section, we evaluate the performance of the finite-domain constraint solver.

We compared the performance of B-Prolog version 6.7 (BP)\footnote{Available from www.probp.com.} with three other CLP(FD) systems: ECLiPSe 5.8 \#77 (EP), GNU-Prolog version 1.2.16 (GP), and SICStus 3.12.0 (SP). There are two solvers available in BP: one is called BP-AC which adopts the hybrid algorithm presented in this paper for equality constraints and the other called BP-IC which maintains only interval consistency for equality constraints. BP-AC is the default solver.\footnote{To switch to BP-IC, set the Prolog flag {\tt constr\_consistency} to {\tt int}}. The linear-space propagator is used for {\tt all\_distinct} in both solvers.

\begin{table}
\begin{small}
\begin{center}
\caption{\label{tab:time}Comparison of CPU times.}
\begin{tabular}{crrrrr|rr} \hline
{\tt Program} &  \multicolumn{5}{c|}{\bf XP} & \multicolumn{2}{c}{\bf Linux} \\ 
\cline{2-8} & {\tt BP-IC} & {\tt BP-AC} & {\tt GP} & {\tt EP} & {\tt SP} & {\tt BP-IC} & {\tt GP} \\ \hline
alpha        &1 &0.82 &1.02 &7.23 &3.17 & 1 & 0.77\\
bridge          &1 &0.94 &1.20 &3.60 &3.57 & 1 & 0.92\\ 
cars    &1 &1.00 &1.67 &7.03 &4.67 & 1 & 1.60 \\
color   &1 &1.14 &1.00 &6.29 &3.01 & 1 & 1.25\\
eq10    &1 &0.98 &3.77 &4.77 &4.92 & 1 & 3.78\\
eq20    &1 &1.06 &2.00 &4.23 &3.34 & 1 & 1.96\\
magic3          &1 &1.38 &1.98 &8.44 &4.41 & 1 & 1.52 \\
magic4          &1 &1.18 &1.96 &8.45 &6.27 & 1 & 1.57 \\
olympic         &1 &0.75 &2.25 &11.25 &4.75 & 1 & 1.43 \\
queens(25)        &1 &1.01 &0.43 &4.24 &5.03 & 1 & 0.43 \\
sendmoney       &1 &1.09 &3.65 &6.74 &7.78 & 1 & 2.62 \\
sudoku81        &1 &1.00 &2.28 &6.67 &6.18 & 1 & 1.36 \\
zebra   &1 &1.13 &2.33 &7.86 &9.61 & 1 & 1.77 \\ \hline
$<$arithmetic mean$>$   &1 &1.04 &1.96 &6.68 &5.13 & 1 & 1.61\\
$<$geometric mean$>$   &1 &1.03 &1.72 &6.36 & 4.84 &1 & 1.42 \\ \hline
\end{tabular}
\end{center}
\end{small}
\end{table}

Table \ref{tab:time} shows the CPU times taken by the four solvers to run a set of benchmarks,\footnote{Available from probp.com/bench.tar.gz.} assuming the time taken by BP-IC be 1. Most of the benchmarks have been widely used by other authors to compare CLP(FD) systems \cite{Carlsson97,COD96,HEN89}. Three new programs were added by the author into the set: {\tt color} is a program that colors a map with 110 regions; {\tt olympic} is a puzzle taken from a Mathematics Olympic game for elementary students; and {\tt sudoku81} is a program for solving a puzzle. The left-to-right labeling strategy is used to instantiate variables in all the benchmarks. The CPU times were measured on a 1.7GHz CPU running Windows XP.  Each program was run at least 10 times and the average was taken. For some programs, execution was repeated up to 1000 times to obtain a stable average. Garbage collection was disabled. GP has a native code compiler for Linux. The comparison with GP was also conducted on Linux.

The BP solvers compare favorably with GP and are significantly faster than EP and SP. EP is the slowest among the compared solvers, probably because of the overhead of supporting priority-based scheduling \cite{Wallace04}. BP outperforms GP remarkably for programs that contain non-binary equality constraints, such as {\tt eq10}, {\tt eq20}, and {\tt sendmoney}. This result reveals that the disadvantages of splitting n-ary constraints into indexicals outweigh the advantages. On the other hand, GP is more than twice as fast as BP for {\tt queens(25)}. The high speed for {\tt queens} may be attributed to an optimization technique adopted in GP that combines indexicals. If the propagators for the disequality constraints were combined for the program in BP,\footnote{For the three disequality constraints $X \neq Y$, $X \neq Y+N$, and $X \neq Y-N$, we can use one propagator rather than three to handle the $ins(X)$ and $ins(Y)$ events.} the speed would be doubled. 

Comparing BP-AC and BP-IC reveals that the hybrid algorithm is effective for {\tt alpha} and {\tt olympic} only. The BP-AC solver is adopted as the default one since for some programs, such as the queens program given in \cite{PUG95},\footnote{This program is not included in the benchmark set since it requires support of negative integer domains and thus cannot run on GP.} BP-AC is exponentially faster than BP-IC. BP-AC is slightly slower than BP-IC for some of the programs. In general, this happens for programs for which the efforts to reduce domains do not pay off. 

\begin{table}
\begin{center}
\caption{\label{tab:back}Comparison of numbers of backtracks.}
\begin{tabular}{crrr} \hline\hline
{\tt Program} &  {\tt BP-AC} & {\tt BP-IC} &  {\tt GP} \\ \hline 
alpha              & {\it 4605} & 8440  & 8440  \\ 
bridge             & 0    &     0 &   0   \\ 
cars               & 53    &   53 &  {\it 34} \\ 
color              & 560   &  560 & 560   \\ 
eq10               & 49    &   49 &  49   \\ 
eq20               & 49    &   49 &  49   \\ 
magic3             & 2     &   2 &    2   \\ 
magic4             & 18    &  18 &   18   \\ 
olympic            & {\it 36}    &  50 &   50   \\ 
queens(25)         & 7255  & 7255 & 7255  \\ 
sendmoney          & 2    &    2 &    2  \\ 
sudoku81           & 0    &    0 &    0  \\ 
zebra              & 2    &    2 &    2  \\ \hline \hline
\end{tabular}
\end{center}
\end{table}

Table \ref{tab:back} gives the numbers of backtracks performed by the three solvers. {\tt BP-AC} makes the same number of backtracks as {\tt BP-IC} except for {\tt alpha} and {\tt olympic},  and {\tt GP} makes fewer backtracks than {\tt BP-AC} for {\tt cars}. Basically, the three solvers explore the same search trees for most of the programs. Therefore, the comparison results shown in Table \ref{tab:time} reflect the real performance of the solvers.

GP and BP are quite different. In GP constraints are compiled into indexicals defined in C while in BP constraints are compiled into propagators defined in action rules. Although the GP Prolog engine may not have much impact on the performance of constraint programs, the BP engine does have a great impact since all the propagators are defined in action rules. One evidence for this observation is that the BP constraint solver becomes 20-30 percent faster after the main switch statement in the emulator is changed to a jump table. A further speed-up is expected if a native code compiler is employed.

There are other factors that affect the performance of a solver, such as domain representation, interaction with other solvers, and garbage collection\cite{Wallace04}. GP supports only finite domains of positive integers, while BP supports not only finite integer domains but also trees and finite domains of ground terms and sets \cite{Zhou02}. In BP, integer domains are represented as described in Subsection~\ref{sec_domain_variables}. BP adopts a sound and complete arithmetic that guarantees that solutions found are correct and no solution is lost. When excluding an inner element from a large interval domain, the system generates a disequality constraint rather than brutally changes the interval into a bit vector as is done in GP.\footnote{Generating a disequality constraint is less efficient than changing an interval into a bit vector since the disequality constraint needs to be checked each time the variable is instantiated.} BP has a garbage collector that collects garbage on the heap and the control stack, but GP does not support garbage collection yet. Garbage collection may suppress some optimization techniques.

\section{Related Work}
CLP(FD) systems have undergone an evolution process, from closed to open and from low level to high level. Several constructs have been proposed to facilitate the implementation of constraint propagators. Examples include attributed variables \cite{HOL92}, indexicals \cite{COD96}, extended indexicals called projection constraints \cite{SID96}, delay clauses \cite{Meier94,Zhou98}, and constraint handling rules \cite{FRU98}. An action rule is an extension of a delay clause that allows for the descriptions of not only delay conditions on subgoals but also activating events and actions. This section compares AR with these constructs introduced into Constraint Logic Programming. Constructs introduced into other languages such as ILOG \cite{PUG95} and Oz \cite{SCHU00} are not compared.

AR is more powerful and flexible than indexicals. We have described in this paper several propagation algorithms in AR, some of which cannot be encoded in indexicals (e.g., the hybrid algorithm for n-ary constraints) and some of which cannot be implemented as efficiently (e.g., arc consistency for binary constraints). Consider the following indexical taken from \cite{Carlsson97},
\[X\ in\ dom(Y)+C\]
which maintains arc consistency for the constraint $X = Y+C$ w.r.t. $X$. Whenever an element $y$ is excluded from the domain of $Y$, the indexical is activated. Because the indexical does not know what the excluded element is, it has to go through the domain elements of $Y$ in the worst case to locate a possible no-good value in the domain of $X$. In contrast, in the propagator implemented in AR, the propagator knows exactly what element is excluded from the domain of $Y$ and thus can compute the counterpart in the domain of $X$ in constant time. 

Compiling constraints into indexicals enables the use of more specialized propagators and restricts propagation to within a small number of constraints \cite{COD96}. Nevertheless, this approach has to introduce new temporary variables and lower the granularity of propagators. Our experiment reveals that B-Prolog outperforms GNU-Prolog for almost all the benchmarks that contain non-binary constraints. This result reveals that the disadvantages of splitting constraints outweigh the advantages. Similar observations have been made independently in \cite{Carlsson97,HAR98,Zhou98}. In \cite{HAR98}, the same two-phase algorithm is used to reduce domains of linear constraints.

Attributed variables \cite{HOL92} are variables with attached attributes each of which has a list of handlers. Touching an attribute triggers the corresponding list of handlers. In order to make context switching swift for handlers, systems such as ECLiPSe treats handlers as demons rather than as normal subgoals. A demon is different from a normal subgoal in that it does not disappear after execution but instead waits for another activation. In this sense, agents in our system are similar to demons. Nevertheless, an agent can be activated by different kinds of events and an agent may take different actions depending on the conditions. An agent can be defined by multiple action rules and the rules are compiled into a tree by the compiler such that shared tests are combined and conditions that failed once need not be tested again. SICStus \cite{Carlsson97} provides interfaces for implementing propagators and also some sort of delay construct similar to attributed variables that triggers propagators when events are posted. 

AR is an extension of our early delay construct proposed in \cite{Zhou98} that allows for the event $dom(X,E)$ and user-defined events. The support of the event $dom(X,E)$ is essential for implementing arc consistency algorithms and also propagators for set constraints \cite{Zhou02}. Our  delay construct is an extension of Meier's delay clause construct \cite{Meier94} that allows for not only delay conditions but also events and actions. In Meier's delay clause, events are implicitly extracted from delay conditions and a delayed subgoal never takes actions as long as the delay condition is satisfied. In retrospect, all these constructs were inspired by early work by Colmerauer and Naish \cite{Colmerauer84,Naish85}.

Other rule-based languages have been designed for implementing constraint propagators. CHR resembles a production system. In CHR, the left-hand side of a rule specifies a pattern of constraints in the constraint store and the right-hand side specifies new constraints to replace those on the left-hand side or to be added into the store. It should be possible to implement in CHR all the propagation algorithms described in this paper provided certain built-ins are added. If events are treated as constraints, then an action rule can be translated into a CHR rule. Treating events as constraints, however, can hardly achieve the same performance. Events are removed automatically after all the agents that are waiting for them are activated. In CHR, there must be rules to remove the events explicitly. The left-hand sides of CHR rules can have multiple constraint patterns. Therefore, it is impossible in general to translate a CHR rule into action rules straightforwardly. It is not clear whether or not it is possible to simulate CHR rules in action rules and how if the answer is yes. It would be an interesting direction to explore in the future.

\section{Conclusion}
There is a need for an implementation language for constraint propagators that is expressive enough and can be implemented efficiently. This paper has presented such a language called AR. The expressiveness of the language is illustrated though several examples that cannot be implemented in indexicals: the propagator for maintaining arc consistency of binary equality constraints; a weak arc-consistency propagator for the {\tt all\_distinct} constraint; and a hybrid algorithm for non-binary equality constraints that combines interval and arc consistency ones. The efficiency is evaluated through benchmarking. For a set of widely used benchmarks, our solver implemented in B-Prolog is significantly faster than that of GNU-Prolog, one of the fastest finite-domain constraint solvers available now. 

The results are encouraging and promising since our solver is implemented in a high-level language and B-Prolog is an emulator-based system which provides more facilities than GNU-Prolog such as garbage collection and constraint solving over other domains. The high-performance of our solver stems from the following facts. Firstly, only one propagator is generated for each non-binary equality constraint that maintains interval consistency. Our solver performs especially well for the benchmarks that contain non-binary equality constraints. This reveals that compiling non-binary equality constraints into indexicals has more cons than pros. Secondly, the hybrid algorithm adopted in our solver is a good compromise between the need to achieve high-level consistency to cut search spaces and the need to reduce the cost. The cost of achieving arc consistency for binary constraints is relatively small, but the effect can be very big for certain programs. Thirdly, our solver employs optimization techniques that reduce redundant activations of propagators.

Our solver can be improved further in the following aspects: (1) develop new optimization techniques for further avoiding redundant activations of propagators; and (2) implement consistency algorithms beyond interval and arc consistency such as path consistency and Regin's algorithm for {\tt all\_distinct}.

\section*{Acknowledgement}
The author would like to thank Kazunori Ueda and the referees for very detailed and helpful comments on the presentation. This research is supported in part by RF-CUNY and CUNY Software Institute.

\bibliography{zhou.bbl}
\bibliographystyle{acmtrans}
\end{document}